\documentclass[a4paper,11pt]{article}
\usepackage[dvips]{graphicx}
\usepackage{amssymb}
\usepackage{amsmath}

\usepackage{cite}

\input arrow

\begin{document}

\title{Localized Fermions on Domain Walls and Extended Supersymmetric Quantum Mechanics}
\author{V. K. Oikonomou\thanks{
voiko@physics.auth.gr}\\
Max Planck Institute for Mathematics in the Sciences\\
Inselstrasse 22, 04103 Leipzig, Germany} \maketitle

\begin{abstract}
We study fermionic fields localized on topologically unstable domain walls bounded by strings in a grand unified theory theoretical framework. Particularly, we found that the localized fermionic degrees of freedom, which are up and down quarks as well as charged leptons, are connected to three independent $N=2$, $d=1$ supersymmetric quantum mechanics algebras. As we demonstrate, these algebras can be combined to form higher order representations of $N=2$, $d=1$ supersymmetry. Due to the uniform coupling of the domain wall solutions to the down-quarks and leptons, we also show that a higher order $N=2$, $d=1$ representation of the down-quark--lepton system is invariant under a duality transformation between the couplings. In addition, the two $N=2$, $d=1$ supersymmetries of the down-quark--lepton system, combine at the coupling unification scale to form an $N=4$, $d=1$ supersymmetry. Furthermore, we present the various extra geometric and algebraic attributes that the fermionic systems acquire, owing to the underlying $N=2$, $d=1$ algebras.
\end{abstract}

\section*{Introduction}

Realistic grand unified theories give rise to various extended topological structures such as monopoles, strings, domains walls and various interesting combinations of these \cite{vilenkin,lazaridesvasiko,lazarides1,lazarides2so10,domainsusypaper,dejan}. Cosmic strings are rather plausible ingredients for cosmological applications, since if they are associated with a symmetry breaking scale of the order $\sim 10^6$GeV, they can serve as a source of primordial density fluctuations in the context of inflationary cosmology. Of course, these have to disappear before they dominate energetically the energy density of the universe. However, monopoles and topologically stable domain walls are rather cosmologically unacceptable. In the case of monopoles, there exists the problem of overproduction which naturally appears in most of the grand unified theories. Domains walls on the other hand, associated with any reasonable mass scale, are a serious problem for cosmology, since they can lead to inconsistencies, with regard to the observed universe. Hence, only unstable domain walls are acceptable for cosmological reasons, and on this kind of domain walls is the focus in this paper. Particularly, we shall be interested for domain walls bounded by cosmic strings \cite{lazaridesvasiko,lazarides1,lazarides2so10}, which can naturally appear when certain grand unified theories are spontaneously broken. The occurrence of domain walls bounded by strings does not necessarily entail catastrophic consequences for cosmology. This kind of domain walls is topologically unstable and it is likely that they disappear before dominating the expansion of the universe. In addition, these domain walls are locally stable and lose energy through their interaction with the surrounding medium. A domain wall is a topological defect that can be created when spontaneous symmetry breaking occurs for a discrete symmetry of a quantum gauged system. Note that, the discrete broken symmetry is not part of the gauge symmetry of the gauge theory. In such a case, the total vacuum manifold consists of various distinct vacuum states, with the field related to the spontaneously broken symmetry taking one of these different vacuum states \cite{lazaridesvasiko,lazarides1,lazarides2so10}. These different vacuum regions are spatially separated by domain walls and the aforementioned field interpolates between these distinct vacuum states. A domain wall can be superconducting in the sense that it can support localized fermionic zero modes which propagate along the wall, and the wall can acquire charge and electric current, with the latter two giving rise to long range electromagnetic interactions \cite{lazaridesvasiko,lazarides1,lazarides2so10,domainsusypaper}. In addition, it has been speculated that the superconducting defects can play some role in generating primordial magnetic fields. 

In this paper we shall study some particular grand unified theories models that give rise to topologically unstable superconducting domain walls. Particularly, we shall focus on the localized fermionic zero modes on the domain wall and we shall reveal an $N=2$ Supersymmetric Quantum Mechanics (hereafter SUSY QM) algebra,  underlying every fermionic family of the model. Moreover, we shall demonstrate that the various different SUSY QM algebras can combine in certain situations to form a larger reducible representation of $N=2$, $d=1$ SUSY QM, or central charge extended $N=2$ SUSY QM algebras. Interestingly enough, under some specific conditions, we shall evince that some of the algebras are enhanced to $N=4$ SUSY QM with central charge. In addition, we shall provide some geometric and algebraic attributes that are a direct consequence of the SUSY QM algebra, to each of the fermionic systems under study. 

The existence of a SUSY QM algebra underlying such a fermionic system is rather interesting. Supersymmetric Quantum Mechanics is a field of research which is by itself interesting and does not serve as being just an one dimensional theoretical tool for testing higher dimensional quantum field theories. Indeed, it was realized that SUSY QM gives insight into the factorization method, with the latter being the first method used to categorize the analytically solvable potential problems. For very informative reviews see \cite{reviewsusyqm,susyqm}. The applications of SUSY QM are numerous, for example various features of extended supersymmetries and harmonic superspaces are studied in \cite{extendedsusy,ivanov}, while some applications of SUSY QM in scattering can be found in \cite{susyqmscatter}. Various applications in quantum mechanical systems can be found in \cite{various,susyqminquantumsystems}. Interesting features of supersymmetry breaking can be found in \cite{susybreaking}, while some geometrical applications of SUSY QM methods can be found in \cite{diffgeomsusyduyalities}. In addition some applications to extended SUSY QM algebras can be found in \cite{plu1,plu2,plu3,plu4}. Although SUSY QM and global spacetime supersymmetry can be related for some theories \cite{ivanov}, in general these are completely different concepts. We shall try to make this clear by using some convincing counter-examples. Particularly, a global $N=1$ supersymmetry model that gives rise to superconducting domain walls with an $N=2$ SUSY QM algebra underlying this system too.

This paper is organized as follows: In section 1 we present the grand unified models we shall use and derive the fermionic equations of motion which will be the starting point of our analysis. We shall demonstrate that an $N=2$, $d=1$ SUSY QM algebra underlies each fermionic system separately. In addition, we shall evince that the SUSY QM algebras can be central charge extended and also can combine in certain cases to form higher dimensional reducible representations of $N=2$ or even, $N=4$, $d=1$ irreducible representations with central charge. Moreover, a model of superconducting domain walls with global $N=1$ supersymmetry is studied, and we still find an underlying $N=2$ SUSY QM. In the end of section 1, we explain the difference of SUSY QM and global spacetime supersymmetry. In section 2, we present the implications of SUSY QM on the fermionic systems. Particularly, we shall demonstrate that for each fermionic system, there exists an underlying global $U(1)$ symmetry and a spin complex structure. In addition, we describe the local geometrical implications of the SUSY QM on the space of the fermionic sections of the corresponding fibre bundle. The conclusions follow in the end of the article.

\section{Superconducting Domain Walls and SUSY QM}

\subsection{Superconducting Domain Walls Essentials}

In this section, the focus is on the brief description of the grand unified theory that admits superconducting domain wall solutions. For details on the issues that will be presented, see \cite{lazaridesvasiko,lazarides2so10,lazarides1}. We shall consider a symmetry breaking pattern of an $SO(10)$ grand unified theory, in which case domain walls bounded by cosmic strings occur. Particularly, we shall be interested in the breaking pattern which occurs through the $126$-Higgs representation. The real grand unification symmetry of the quantum system is not the $SO(10)$ group, but the $Spin(10)$, owing to the fact that the fermions are in the $16$-representation, which is the fundamental spinor representation of $SO(10)$. The symmetry breaking pattern is the following:
\begin{equation}\label{symmetrybreakpatt}
Spin(10)\xrightarrow{54,M_x}H_1\xrightarrow{126,M_R}H_2\xrightarrow{10,M_w}SU(3)_c\times U(1)_{em}
\end{equation}
In the above, the mass scales $M_x,M_R$ are of the order $M_x\sim 10^{15}$GeV and $M_R\sim 10^{13}$GeV. In addition, the subgroups $H_1$ and $H_2$ are equal to:
\begin{align}\label{subgroupsbrek}
& H_1\sim Spin(6)\times Spin(4) \\ \notag & H_2\sim SU(3)_c\times SU(2)_L\times U(1)
\end{align}
Actually, the group $H_1$ is isomorphic to the Pati-Salam subgroup, that is, $H_1\sim SU(4)\times SU(2)_L\times SU(2)_R$. Since the subgroup $H_1$ is disconnected, the fundamental homotopy group is $\pi (H_1)=Z_2$. This $Z_2$ is generated by the disconnected piece of $H_1$, which we denote $C$. Note that $H_1=H_0'\times C$, with the subgroup $H_0'$ being equal to:
\begin{equation}\label{hosubgroup}
H_0'=(Spin(6)\times Spin(4))/Z_2
\end{equation}
At the stage in which the symmetry of the system is $H_1$, topologically stable strings occur. At the second symmetry breaking stage, the discrete symmetry $C$ is broken and consequently domain walls appear, separating regions with opposite values of $C$. The domain walls terminate on the $H_1$-phase strings, and these are not completely stable, because the can decay into holes bounded by string loops.

The scalar field that will give rise to domain walls is denoted by $\phi$, which can take two vacuum expectation values, namely $\langle \phi \rangle=\phi_u$ and $\langle \phi \rangle=\phi_d$. Going through the wall, we can connect the corresponding $-\langle \phi \rangle$ value to the $\langle \phi \rangle$ value, with the aid of a kink configuration. What we are mainly interested for, is the existence of fermionic localized degrees of freedom on the wall. It is exactly the existence of these massless localized zero modes that renders the domain walls superconducting. On the wall we have both left handed fields $\psi_i=(u_L,d_L,e_L)$ and right handed fields $\chi_i=(u_R,d_R,e_R)$, with $(u,d,e)$ denoting the up and down quarks and the electron field respectively. Note that $\phi_d$ couples to the down quark and the charged leptons, while $\phi_u$ couples to the up-quarks only. This fact has an interesting consequence on the SUSY QM structure of the lepton-down quark system, as we shall see in a later section. The equations of motion for each right-handed and left 
handed fermionic field $(\chi_i,\psi_i)$ are equal to \cite{lazaridesvasiko}:
\begin{align}\label{fer1}
&i\partial_0\psi_{i}-\sigma^i(i\partial_i-A_i)\psi_i-g_i\phi_k^*(y)\chi_i=0
\\ \notag & i\partial_0\chi_{i}+\sigma^i(i\partial_i-A_i)\chi_i-g_i\phi_k^*(y)\psi_i=0
\end{align}
with $\phi_k(y)=\phi_u(y),\phi_d(y)$ depending on which fermion is coupled to the $\phi$ field, and $g_i$ is a real coupling which takes the values $g_i=(g_u,g_d,g_e)$ when $\phi (y)$ is coupled to the up quark, the down quark, and the charged lepton respectively. Assuming an infinite wall in the $x-z$ plane and also that $A_i=0$, the transverse fermionic zero modes in the $y$-direction satisfy the following equation:
\begin{align}\label{fer1}
 -&i\sigma^2\partial_y\psi_i(y)-g_i\phi_k^*(y)\chi_i(y)=0
\\ \notag & i\sigma^2\partial_y\chi_i(y)-g_i\phi_k^*(y)\psi_i(y)=0
\end{align}
As is established in \cite{lazaridesvasiko}, a real kink solution $\phi_k(y)=\phi_k^{kink}(y)$ can always be found, hence we shall assume that $\phi(y)$ is real. Then, the equations of motion for the up-quark, down-quark and charged leptons are:
\begin{align}\label{fer1}
 -&i\sigma^2\partial_y\psi_u(y)-g_u\phi_u^{kink}(y)\chi_u(y)=0
\\ \notag & i\sigma^2\partial_y\chi_u(y)-g_u\phi_u^{kink}(y)\psi_u(y)=0
\\ \notag -& i\sigma^2\partial_y\psi_d(y)-g_d\phi_d^{kink}(y)\chi_d(y)=0
\\ \notag & i\sigma^2\partial_y\chi_d(y)-g_d\phi_d^{kink}(y)\psi_d(y)=0
\\ \notag -& i\sigma^2\partial_y\psi_e(y)-g_e\phi_e^{kink}(y)\chi_e(y)=0
\\ \notag & i\sigma^2\partial_y\chi_e(y)-g_e\phi_e^{kink}(y)\psi_e(y)=0
\end{align}
with $\phi_e^{kink}(y)=\phi_d^{kink}(y)$, since the kink couples in the same way to the down quark and lepton sector. Each set of the above equations (\ref{fer1}) admits localized solutions, which are actually:
\begin{align}\label{soleqnts}
& \psi_u(y)=c_ue^{-g_u\int_0^y\phi_u^{kink}(y)\mathrm{d}y},{\,}{\,}{\,}\chi_u(y)=c_ui\sigma^2e^{-g_u\int_0^y\phi_u^{kink}(y)\mathrm{d}y}
\\ \notag & \psi_d(y)=c_de^{-g_d\int_0^y\phi_d^{kink}(y)\mathrm{d}y},{\,}{\,}{\,}\chi_d(y)=c_di\sigma^2e^{-g_d\int_0^y\phi_d^{kink}(y)\mathrm{d}y}\\ \notag & \psi_e(y)=c_ee^{-g_e\int_0^y\phi_e^{kink}(y)\mathrm{d}y},{\,}{\,}{\,}\chi_e(y)=c_ei\sigma^2e^{-g_e\int_0^y\phi_e^{kink}(y)\mathrm{d}y}
\end{align}
The existence of these transverse localized zero modes is what renders the domain wall superconducting. In the following we focus on one set of fermionic fields, for example the set $(\chi_u(y),\psi_u(y))$, but the results hold for the other two sets of fermionic fields.

\subsection{Localized Fermions and SUSY QM-Zero Central Charge Case}

Consider the first two equations of relation (\ref{fer1}). These can be written in terms of the operator $D_u$:
\begin{equation}\label{susyqmrn567m}
\mathcal{D}_{u}=\left(%
\begin{array}{cc}
-i\sigma^2\partial_y & g_u\phi_u^{kink}(y)
 \\ g_u\phi_u^{kink}(y) & i\sigma^2\partial_y\\
\end{array}%
\right)
\end{equation}
which acts on the vector:
\begin{equation}\label{ait34e1}
|\Psi_{u}\rangle =\left(%
\begin{array}{c}
  \psi_{u}(y) \\
  \chi_u(y) \\
\end{array}%
\right).
\end{equation}
Thereby, the first two equations of equation (\ref{fer1}), can be written in the following form:
\begin{equation}\label{transf}
\mathcal{D}_{u}|\Psi_{u}\rangle=0
\end{equation}
The solutions of equation (\ref{transf}) are actually the normalizable zero mode eigenfunctions of the operator $\mathcal{D}_{u}$. Since there is only one localized normalizable solution for the fermionic fields $(\chi_u(y),\psi_u(y))$, as can be seen from equation (\ref{soleqnts}), we may easily conclude that:
\begin{equation}\label{dimeker}
\mathrm{dim}{\,}\mathrm{ker}\mathcal{D}_{u}=1
\end{equation}
In addition, the adjoint of the operator $\mathcal{D}_{u}$, that is $\mathcal{D}_{u}^{\dag}$, is:
\begin{equation}\label{eqndag}
\mathcal{D}_{u}^{\dag}=\left(%
\begin{array}{cc}
i\sigma^2\partial_y & g_u\phi_u^{kink}(y)
 \\ g_u\phi_u^{kink}(y) & -i\sigma^2\partial_y\\
\end{array}%
\right)
\end{equation}
and acts on the vector:
\begin{equation}\label{ait3hgjhgj4e1}
|\Psi_{u}'\rangle =\left(%
\begin{array}{c}
  \chi_{u}(y) \\
  \psi_{u}(y) \\
\end{array}%
\right).
\end{equation}
The zero mode solutions of the adjoint operator $\mathcal{D}_{u}^{\dag}$, are also the solutions of the first two equations of (\ref{fer1}), and hence are identical to the solutions of the operator $D_u$. Therefore, in this case too, the corresponding kernel of the adjoint operator is:
\begin{equation}\label{dimeke1r11}
\mathrm{dim}{\,}\mathrm{ker}\mathcal{D}_{u}^{\dag}=1
\end{equation}
We took into account only normalizable solutions of (\ref{fer1}), in which case, the operator $\mathcal{D}_{u}$ is Fredholm, a result that can be verified by (\ref{dimeker}) and (\ref{dimeke1r11}). The Fredholm property of the associated operators is an exceptional attribute of the fermionic systems that we study.

\noindent The fermionic system of the fields $(\psi_u(y),\chi_u(y))$ that are localized on the domain wall, has an unbroken $N=2$, $d=1$ supersymmetry with the supercharges of this $N=2$, $d=1$ SUSY algebra being equal to:
\begin{equation}\label{s7}
\mathcal{Q}_{u}=\bigg{(}\begin{array}{ccc}
  0 & \mathcal{D}_{u} \\
  0 & 0  \\
\end{array}\bigg{)},{\,}{\,}{\,}\mathcal{Q}^{\dag}_{u}=\bigg{(}\begin{array}{ccc}
  0 & 0 \\
  \mathcal{D}_{u}^{\dag} & 0  \\
\end{array}\bigg{)}
\end{equation}
In addition, the quantum Hamiltonian of the quantum system is:
\begin{equation}\label{s11}
\mathcal{H}_{u}=\bigg{(}\begin{array}{ccc}
 \mathcal{D}_{u}\mathcal{D}_{u}^{\dag} & 0 \\
  0 & \mathcal{D}_{u}^{\dag}\mathcal{D}_{u}  \\
\end{array}\bigg{)}
\end{equation}
These operators satisfy the one dimensional SUSY QM algebra:
\begin{equation}\label{relationsforsusy}
\{\mathcal{Q}_{u},\mathcal{Q}^{\dag}_{u}\}=\mathcal{H}_{u}{\,}{\,},\mathcal{Q}_{u}^2=0,{\,}{\,}{\mathcal{Q}_{u}^{\dag}}^2=0
\end{equation}
The quantum Hilbert space of the supersymmetric quantum mechanical system, namely $\mathcal{H}$, is rendered $Z_2$-graded by the operator $\mathcal{W}$, which is called the Witten parity, and it is actually an involution operator. This operator commutes with the total Hamiltonian,
\begin{equation}\label{s45}
[\mathcal{W},\mathcal{H}_{u}]=0
\end{equation}
and anti-commutes with the supercharges,
\begin{equation}\label{s5}
\{\mathcal{W},\mathcal{Q}_{u}\}=\{\mathcal{W},\mathcal{Q}_{u}^{\dag}\}=0
\end{equation}
Furthermore, $\mathcal{W}$ satisfies the following identity,
\begin{equation}\label{s6}
\mathcal{W}^{2}=1
\end{equation}
which is a characteristic property of projection operators. It worths demonstrating what kind of projection operator this involution is. The Witten parity $\mathcal{W}$, spans the total Hilbert space into $Z_2$ equivalent subspaces, with the total Hilbert space of the quantum system being written as:
\begin{equation}\label{fgjhil}
\mathcal{H}=\mathcal{H}^+\oplus \mathcal{H}^-
\end{equation}
The vectors belonging to the two subspaces $\mathcal{H}^{\pm}$, are classified to even and odd parity states, according to their Witten parity, that is:
\begin{equation}\label{shoes}
\mathcal{H}^{\pm}=\mathcal{P}^{\pm}\mathcal{H}=\{|\psi\rangle :
\mathcal{W}|\psi\rangle=\pm |\psi\rangle \}
\end{equation}
Moreover, the Hamiltonians corresponding to the $Z_2$ graded spaces are:
\begin{equation}\label{h1}
{\mathcal{H}}_{+}=\mathcal{D}_{u}{\,}\mathcal{D}_{u}^{\dag},{\,}{\,}{\,}{\,}{\,}{\,}{\,}{\mathcal{H}}_{-}=\mathcal{D}_{u}^{\dag}{\,}\mathcal{D}_{u}
\end{equation}
In our case, the operator $\mathcal{W}$, has the following matrix form:
\begin{equation}\label{wittndrf}
\mathcal{W}=\bigg{(}\begin{array}{ccc}
  1 & 0 \\
  0 & -1  \\
\end{array}\bigg{)}
\end{equation}
The eigenstates of the operator $\mathcal{P}^{\pm}$, namely $|\psi^{\pm}\rangle$, satisfy
the following relation:
\begin{equation}\label{fd1}
P^{\pm}|\psi^{\pm}\rangle =\pm |\psi^{\pm}\rangle
\end{equation}
Therefore, we shall call them positive and negative parity eigenstates, with ``parity'' referring to the $P^{\pm}$ operator, which is related to the Witten parity operator. Using the representation (\ref{wittndrf}) for the Witten parity operator, the parity eigenstates can be represented by the vectors,
\begin{equation}\label{phi5}
|\psi^{+}\rangle =\left(%
\begin{array}{c}
  |\phi^{+}\rangle \\
  0 \\
\end{array}%
\right),{\,}{\,}{\,}
|\psi^{-}\rangle =\left(%
\begin{array}{c}
  0 \\
  |\phi^{-}\rangle \\
\end{array}%
\right)
\end{equation}
with $|\phi^{\pm}\rangle$ $\epsilon$ $\mathcal{H}^{\pm}$. It worths writing the fermionic states $(\psi_u(y),\chi_u(y))$ in terms of the SUSY quantum algebra. It easy to identify that:
\begin{equation}\label{fdgdfgh}
|\Psi_{u}\rangle =|\phi^{-}\rangle=\left(%
\begin{array}{c}
  \psi_{u}(y) \\
  \chi_{u}(y) \\
\end{array}%
\right),{\,}{\,}{\,}|\Psi_{u}'\rangle =|\phi^{+}\rangle=\left(%
\begin{array}{c}
  \psi_{u}(y) \\
  \chi_{u}(y) \\
\end{array}%
\right)
\end{equation}
Thereby, the corresponding even and odd parity SUSY quantum states, upon which the Hamiltonian and the supercharges act, are:
\begin{equation}\label{phi5}
|\psi^{+}\rangle =\left(%
\begin{array}{c}
  |\Psi_{u}'\rangle \\
  0 \\
\end{array}%
\right),{\,}{\,}{\,}
|\psi^{-}\rangle =\left(%
\begin{array}{c}
  0 \\
  |\Psi_{u}\rangle \\
\end{array}%
\right)
\end{equation}
In order to see if supersymmetry is unbroken, the Witten index has to be computed for the system at hand. For Fredholm operators, the Witten index is defined to be:
\begin{equation}\label{phil}
\Delta =n_{-}-n_{+}
\end{equation}
with $n_{\pm}$ the number of zero modes of ${\mathcal{H}}_{\pm}$ in the subspace $\mathcal{H}^{\pm}$, with the constraint that these are finitely many. Supersymmetry is considered to be unbroken if the so-called Witten index is a non-zero integer, or in the case $\Delta =0$ with $n_{+}=
n_{-}\neq 0$, in which case the system possesses an unbroken supersymmetry too \cite{reviewsusyqm}. The only case in which supersymmetry is broken is when the Witten index is zero and at the same time $n_{+}=n_{-}=0$. The Witten index is directly connected to the Fredholm index of the operator $\mathcal{D}_{u}$, as follows:
\begin{align}\label{ker1}
&\Delta=\mathrm{dim}{\,}\mathrm{ker}
{\mathcal{H}}_{-}-\mathrm{dim}{\,}\mathrm{ker} {\mathcal{H}}_{+}=
\mathrm{dim}{\,}\mathrm{ker}\mathcal{D}_{u}^{\dag}\mathcal{D}_{u}-\mathrm{dim}{\,}\mathrm{ker}\mathcal{D}_{u}\mathcal{D}_{u}^{\dag}=
\\ \notag & \mathrm{ind} \mathcal{D}_{u} = \mathrm{dim}{\,}\mathrm{ker}
\mathcal{D}_{u}-\mathrm{dim}{\,}\mathrm{ker} \mathcal{D}_{u}^{\dag}
\end{align}
Combining the results of equations (\ref{dimeker}) and (\ref{dimeke1r11}), we can conclude by directly computing the Fredholm index of the operator $D_u$, that the Witten index is equal to:
\begin{equation}\label{fnwitten}
\Delta =0
\end{equation}
and simultaneously $n_{+}=n_{-}=1\neq 0$. Therefore, the fermionic system of the localized fermions $(\psi_u(y),\chi_u(y))$ on the domain wall, has an unbroken underlying $N=2$, $d=1$ supersymmetry. Following the same line of research, it can be easily established that an unbroken $N=2$, $d=1$ SUSY QM underlies each of the other two fermionic systems that are localized on the domain walls, namely $(\chi_d(y),\psi_d(y))$ and $(\chi_e(y),\psi_e(y))$. Indeed, the corresponding supercharges, which we denote $Q_d$ and $Q_e$ are:
\begin{equation}\label{s7supcghagemix}
\mathcal{Q}_{d}=\bigg{(}\begin{array}{ccc}
  0 & \mathcal{D}_{d} \\
  0 & 0  \\
\end{array}\bigg{)},{\,}{\,}{\,}\mathcal{Q}_{e}=\bigg{(}\begin{array}{ccc}
  0 &  \mathcal{D}_{e} \\
 0 & 0  \\
\end{array}\bigg{)}
\end{equation}
Moreover, the quantum Hamiltonians are:
\begin{equation}\label{s11fgghhf}
\mathcal{H}_{d}=\bigg{(}\begin{array}{ccc}
 \mathcal{D}_{d}\mathcal{D}_{d}^{\dag} & 0 \\
  0 & \mathcal{D}_{d}^{\dag}\mathcal{D}_{d}  \\
\end{array}\bigg{)},{\,}{\,}{\,}\mathcal{H}_{e}=\bigg{(}\begin{array}{ccc}
 \mathcal{D}_{e}\mathcal{D}_{e}^{\dag} & 0 \\
  0 & \mathcal{D}_{e}^{\dag}\mathcal{D}_{e}  \\
\end{array}\bigg{)}
\end{equation}
The operators appearing in equations (\ref{s7supcghagemix}) and (\ref{s11fgghhf}) shall be frequently used in the following sections and are equal to:
\begin{equation}\label{twooperators}
\mathcal{D}_{d}=\left(%
\begin{array}{cc}
-i\sigma^2\partial_y & g_d\phi_d^{kink}(y)
 \\ g_d\phi_d^{kink}(y) & i\sigma^2\partial_y\\
\end{array}%
\right),{\,}{\,}{\,}\mathcal{D}_{e}=\left(%
\begin{array}{cc}
-i\sigma^2\partial_y & g_e\phi_e^{kink}(y)
 \\ g_e\phi_e^{kink}(y) & i\sigma^2\partial_y\\
\end{array}%
\right)
\end{equation}

\subsection{Extended Supersymmetric-Higher Representation Algebras}

Let us recapitulate what we have at hand, up to this point. We demonstrated that each fermionic sector of the model we described in the previous section (that is up-quarks, down-quarks, charged leptons), can constitute an unbroken $N=2$ SUSY QM algebra, related to the localized fermionic zero modes on the domain wall. Hence, we have three complex supercharges, that is:
\begin{align}\label{superchargesrecapitulate}
&\mathcal{Q}_d,\mathcal{Q}_d^{\dag},{\,}{\,}{\,}\mathrm{d-quark} \\ & \notag
\mathcal{Q}_u,\mathcal{Q}_u^{\dag},{\,}{\,}{\,}\mathrm{u-quark} \\ & \notag
\mathcal{Q}_e,\mathcal{Q}_e^{\dag},{\,}{\,}{\,}\mathrm{lepton}
\end{align}
The question is whether these SUSY QM algebras can combine in some way to form extended supersymmetries or higher dimensional SUSY QM representations. In addition, is there any symmetry transformation which connects the Hilbert spaces to which these supercharges act? We shall address all these questions in this section.

\subsubsection{Higher Reducible Representation 1}

\noindent Consider the supercharges $(\mathcal{Q}_u,\mathcal{Q}_d)$ and the corresponding operators $({\mathcal{D}}_{u},{\mathcal{D}}_{d})$ for example. The two $N=2$ supersymmetries corresponding to these supercharges can be combined to a higher reducible representation of a single $N=2$, $d=1$ supersymmetry. Indeed, the supercharges of this representation, denoted ${\mathcal{Q}}_{DU}$ and  ${\mathcal{Q}}_{DU}^{\dag}$ are equal to:
\begin{equation}\label{connectirtyrtons}
{\mathcal{Q}}_{DU}= \left ( \begin{array}{cccc}
  0 & 0 & 0 & 0 \\
  {\mathcal{D}}_{d} & 0 & 0 & 0 \\
0 & 0 & 0 & 0 \\
0 & 0 & {\mathcal{D}}_{u}^{\dag} & 0  \\
\end{array} \right),{\,}{\,}{\,}{\,}{\mathcal{Q}}_{DU}^{\dag}= \left ( \begin{array}{cccc}
  0 &  {\mathcal{D}}_{d}^{\dag} & 0 & 0 \\
  0 & 0 & 0 & 0 \\
0 & 0 & 0 & {\mathcal{D}}_{u} \\
0 & 0 & 0 & 0  \\
\end{array} \right)
.\end{equation}
In addition, the Hamiltonian of the combined quantum system $H_T$, reads,
\begin{equation}\label{connections1dtr}
H_{DU}= \left ( \begin{array}{cccc}
  {\mathcal{D}}_{d}^{\dag}{\mathcal{D}}_{d} & 0 & 0 & 0 \\
  0 & {\mathcal{D}}_{d}{\mathcal{D}}_{d}^{\dag} & 0 & 0 \\
0 & 0 & {\mathcal{D}}_{u}{\mathcal{D}}_{u}^{\dag} & 0 \\
0 & 0 & 0 & {\mathcal{D}}_{u}^{\dag}{\mathcal{D}}_{u}  \\
\end{array} \right)
.\end{equation}
The operators (\ref{connectirtyrtons}) and (\ref{connections1dtr}), satisfy the $N=2$, $d=1$ SUSY QM algebra, namely:
\begin{equation}\label{mousikisimagne}
\{ {\mathcal{Q}}_{DU},{\mathcal{Q}}_{DU}^{\dag}\}=H_{DU},{\,}{\,}{\mathcal{Q}}_{DU}^2=0,{\,}{\,}{{\mathcal{Q}}_{DU}^{\dag}}^2=0,{\,}{\,}\{{\mathcal{Q}}_{DU},\mathcal{W}_{DU}\}=0,{\,}{\,}\mathcal{W}_{DU}^2=I,{\,}{\,}[\mathcal{W}_{DU},H_{DU}]=0
.\end{equation}
In this case, the Witten parity operator is equal to:
\begin{equation}\label{wparityopera}
\mathcal{W}_{DU}= \left ( \begin{array}{cccc}
  1 & 0 & 0 & 0 \\
  0 & -1 & 0 & 0 \\
0 & 0 & 1 & 0 \\
0 & 0 & 0 & -1  \\
\end{array} \right)
.\end{equation}
In addition to this representation, we can form equivalent higher dimensional representations for the combined $N=2$, $d=1$ algebra, by making use of the following substitutions:
\begin{equation}\label{setof transformations}
\mathrm{Set}{\,}{\,}{\,}A:{\,}
\begin{array}{c}
 {\mathcal{D}}_{d}\rightarrow {\mathcal{D}}_{d}^{\dag} \\
  {\mathcal{D}}_{u}^{\dag}\rightarrow {\mathcal{D}}_{u} \\
\end{array},{\,}{\,}{\,}\mathrm{Set}{\,}{\,}{\,}B:{\,}
\begin{array}{c}
 {\mathcal{D}}_{d}\rightarrow {\mathcal{D}}_{u}^{\dag} \\
  {\mathcal{D}}_{u}^{\dag}\rightarrow {\mathcal{D}}_{d} \\
\end{array},{\,}{\,}{\,}\mathrm{Set}{\,}{\,}{\,}C:{\,}
\begin{array}{c}
 {\mathcal{D}}_{d}\rightarrow {\mathcal{D}}_{u} \\
  {\mathcal{D}}_{u}^{\dag}\rightarrow {\mathcal{D}}_{d}^{\dag} \\
\end{array}
.\end{equation}
Moreover, another higher order reducible representation of the $N=2$ SUSY QM algebra, equivalent to the one of relation (\ref{connectirtyrtons}), is given by: 
\begin{equation}\label{connectirtyfhfghrtons}
{\mathcal{Q}}_{DU}= \left ( \begin{array}{cccc}
  0 & 0 & 0 & 0 \\
  0 & 0 & 0 & 0 \\
{\mathcal{D}}_{d} & 0 & 0 & 0 \\
0 & {\mathcal{D}}_{u}^{\dag} & 0 & 0  \\
\end{array} \right),{\,}{\,}{\,}{\,}{\mathcal{Q}}_{DU}^{\dag}= \left ( \begin{array}{cccc}
  0 & 0 & {\mathcal{D}}_{d}^{\dag} & 0 \\
  0 & 0 & 0 & {\mathcal{D}}_{u} \\
0 & 0 & 0 & 0 \\
0 & 0 & 0 & 0  \\
\end{array} \right)
.\end{equation}
with the Hamiltonian being,
\begin{equation}\label{connectihgghdhtons1dtr}
H_{DU}= \left ( \begin{array}{cccc}
  {\mathcal{D}}_{d}^{\dag}{\mathcal{D}}_{d} & 0 & 0 & 0 \\
  0 & {\mathcal{D}}_{u}^{\dag}{\mathcal{D}}_{u} & 0 & 0 \\
0 & 0 & {\mathcal{D}}_{u}{\mathcal{D}}_{u}^{\dag} & 0 \\
0 & 0 & 0 & {\mathcal{D}}_{d}{\mathcal{D}}_{d}^{\dag}  \\
\end{array} \right)
.\end{equation}
Obviously, similar considerations can be done for any other pair of fermions, that is for the remaining pairs $(u,e)$ and $(d,e)$, by the appropriate replacements of the corresponding operators, but we omit these for brevity.

\subsubsection{Higher Reducible Representation 2}

In the previous case, we made use of the operators $\mathcal{D}_i$, $i=(u,d,e)$, and formed combined higher order representations of the SUSY QM algebras corresponding to the three fermionic sectors. In this subsection we shall make use of the supercharges directly and form higher order representations. We take for example the $(e,d)$ sector and using the supercharges $(\mathcal{Q}_e,\mathcal{Q}_d)$ we can form the following supercharges:
\begin{equation}\label{s7supcghagemixdesec}
\mathcal{Q}_{de}=\bigg{(}\begin{array}{ccc}
  0 & \mathcal{Q}_{e} \\
  \mathcal{Q}_{d} & 0  \\
\end{array}\bigg{)},{\,}{\,}{\,}\mathcal{Q}_{de}^{\dag}=\bigg{(}\begin{array}{ccc}
  0 &  \mathcal{Q}_{d}^{\dag} \\
 \mathcal{Q}_{e}^{\dag} & 0  \\
\end{array}\bigg{)}
\end{equation}
and the corresponding quantum Hamiltonian:
\begin{equation}\label{s11fgghhfdesec}
\mathcal{H}_{de}=\bigg{(}\begin{array}{ccc}
 \mathcal{Q}_{e}\mathcal{Q}_{e}^{\dag}+ \mathcal{Q}_{d}^{\dag}\mathcal{Q}_{d}& 0 \\
  0 & \mathcal{Q}_{e}^{\dag}\mathcal{Q}_{e}+ \mathcal{Q}_{d}\mathcal{Q}_{d}^{\dag}  \\
\end{array}\bigg{)}
\end{equation}
The supercharges and the Hamiltonian satisfy the following $N=2$, $d=1$ algebra:
\begin{equation}\label{relationsforsusyrep2new}
\{ {\mathcal{Q}}_{de},{\mathcal{Q}}_{de}^{\dag}\}=H_{de},{\,}{\,}{\mathcal{Q}}_{de}^2=0,{\,}{\,}{{\mathcal{Q}}_{de}^{\dag}}^2=0,{\,}{\,}\{{\mathcal{Q}}_{de},\mathcal{W}_{de}\}=0,{\,}{\,}\mathcal{W}_{de}^2=I,{\,}{\,}[\mathcal{W}_{de},H_{de}]=0
\end{equation}
We can construct similar algebras between the sectors of the down-up quark $(u,d)$ and the up quark-lepton $(u,e)$, but since these are equivalent to the one corresponding to relations (\ref{s7supcghagemixdesec}) and (\ref{s11fgghhfdesec}), we omit them. It is interesting to search for transformations or even dualities between the SUSY QM algebras of the fermionic sectors. Since the Higgs field $\phi$ couples in the same way between the down quark and lepton sector, it follows naturally to find if a duality structure exists for that system. Consider for example the transformation $g_{e}\leftrightarrow g_{d}$. Performing this transformation, the quantum system described by the algebra (\ref{s7supcghagemixdesec}), is not invariant under the transformation $g_{e}\leftrightarrow g_{d}$. Nevertheless, we can construct a higher dimensional representation of an $N=2$ SUSY QM algebra between the $(d,e)$ sector, in which case the new $N=2$ algebra is invariant under this transformation.

\subsubsection{Higher Reducible Supersymmetric Representations and Dualities Between Sectors}

Consider the $(d,e)$ sector quantum subsystem. As we demonstrated in the previous sections, in each sector of the $(d,e)$ system, underlies an unbroken $N=2$ SUSY QM algebra with supercharges $(\mathcal{Q}_{e},\mathcal{Q}_{d})$. Using these supercharges we can construct a reducible higher dimensional $N=2$ SUSY QM algebra, with the property that this algebra is invariant under the duality transformation $g_{e}\leftrightarrow g_{d}$. Indeed, the supercharges  of this algebra are:
\begin{equation}\label{s7supcghagemixdesecnew}
\mathcal{Q}_{D}=\bigg{(}\begin{array}{ccc}
  0 & 0 \\
  \mathcal{Q}_{d}+\mathcal{Q}_{e} & 0  \\
\end{array}\bigg{)},{\,}{\,}{\,}\mathcal{Q}_{D}^{\dag}=\bigg{(}\begin{array}{ccc}
  0 &  \mathcal{Q}_{d}^{\dag}+\mathcal{Q}_{e}^{\dag} \\
 0 & 0  \\
\end{array}\bigg{)}
\end{equation}
and the Hamiltonian is:
\begin{equation}\label{s11fgghhfdesecnew}
\mathcal{H}_{D}=\bigg{(}\begin{array}{ccc}
 \mathcal{Q}_{e}^{\dag}\mathcal{Q}_{e}+ \mathcal{Q}_{e}^{\dag}\mathcal{Q}_{d}+\mathcal{Q}_{d}^{\dag}\mathcal{Q}_{e}+\mathcal{Q}_{d}^{\dag}\mathcal{Q}_{d}& 0 \\
  0 & \mathcal{Q}_{e}\mathcal{Q}_{e}^{\dag}+ \mathcal{Q}_{e}\mathcal{Q}_{d}^{\dag}+\mathcal{Q}_{d}\mathcal{Q}_{e}^{\dag}+\mathcal{Q}_{d}\mathcal{Q}_{d}^{\dag}  \\
\end{array}\bigg{)}
\end{equation}
Again these three elements satisfy the $N=2$, $d=1$ SUSY QM algebra,
\begin{equation}\label{relationsforsusyrep2newrep2}
\{ {\mathcal{Q}}_{D},{\mathcal{Q}}_{D}^{\dag}\}=H_{D},{\,}{\,}{\mathcal{Q}}_{D}^2=0,{\,}{\,}{{\mathcal{Q}}_{D}^{\dag}}^2=0,{\,}{\,}\{{\mathcal{Q}}_{D},\mathcal{W}_{D}\}=0,{\,}{\,}\mathcal{W}_{D}^2=I,{\,}{\,}[\mathcal{W}_{D},H_{D}]=0
\end{equation}
Performing the transformation $g_{e}\leftrightarrow g_{d}$, the Hamiltonian is invariant under this transformation. In addition, the commutation and anti-commutation relations (\ref{relationsforsusyrep2newrep2}) remain invariant. Notice the transformation $g_{e}\leftrightarrow g_{d}$ is equivalent with the transformation $\mathcal{Q}_{e}\leftrightarrow \mathcal{Q}_{d}$. 

Apart from the $(d,e)$ system, we can construct a similar higher representation for the $(u,d)$ system, with the replacement $\mathcal{Q}_{e}\leftrightarrow \mathcal{Q}_{u}$ in relations (\ref{s7supcghagemixdesecnew}) and (\ref{s11fgghhfdesecnew}). This new quantum algebra is invariant under the simultaneous transformations:
\begin{equation}\label{newtransf}
g_{d}\leftrightarrow g_{u},{\,}{\,}{\,}\phi_d^{kink}\leftrightarrow \phi_u^{kink}
\end{equation}

\subsection{Localized Fermions and SUSY QM-Central Charge Case}

Each $N=2$ SUSY QM algebra that underlies the fermionic sectors $(u,d,e)$ can be enriched with a real supercharge. Take for example the up-quark SUSY QM algebra. We can extend this algebra to include a real central charge $Z$, in the following way:
\begin{equation}\label{s7supcghagemixcentch}
\mathcal{Q}_{Z}=\bigg{(}\begin{array}{ccc}
  -\eta & 0 \\
  \mathcal{D}_{u} & \eta  \\
\end{array}\bigg{)},{\,}{\,}{\,}\mathcal{Q}_{Z}=\bigg{(}\begin{array}{ccc}
  -\eta &  \mathcal{D}_{u}^{\dag} \\
 0 & \eta \\
\end{array}\bigg{)}
\end{equation}
with ''$\eta$'' some arbitrary $2\times 2$ real matrix. The Hamiltonian of the system in this case is:
\begin{equation}\label{s11fgghhfcentcharge}
\mathcal{H}_{Z}=\bigg{(}\begin{array}{ccc}
 \mathcal{D}_{u}\mathcal{D}_{u}^{\dag}+2\eta^2 & 0 \\
  0 & \mathcal{D}_{u}^{\dag}\mathcal{D}_{u}+2\eta^2  \\
\end{array}\bigg{)}
\end{equation}
The real central charge of the centrally extended $N=2$ SUSY QM algebra is:
\begin{equation}\label{centrextendcc}
Z=\bigg{(}\begin{array}{ccc}
  2\eta^2 & 0 \\
  0 & 2\eta^2  \\
\end{array}\bigg{)}.
\end{equation}
The supercharges and the Hamiltonian satisfy the following commutation relations:
\begin{equation}\label{relationsforsusycc}
\{\mathcal{Q}_{Z},\mathcal{Q}^{\dag}_{Z}\}=\mathcal{H}_{Z}{\,}{\,},\{\mathcal{Q}_{Z},\mathcal{Q}_{Z}\}=Z,{\,}{\,}\{\mathcal{Q}^{\dag}_{Z},\mathcal{Q}^{\dag}_{Z}\}=Z,{\,}{\,}{\,}[\mathcal{H}_{Z},\mathcal{Q}_{Z}]=[\mathcal{H}_{Z},\mathcal{Q}_{Z}^{\dag}]=0
\end{equation}
The difference between the non-zero central charge and zero central charge case, is that in the former case, the supercharges no longer map the parity-even to parity-odd states, regarding the non-zero modes. In the particular case when the matrix $2\eta^{2}$ is an odd compact matrix, then the index of the operator $\mathcal{D}_{u}$ remains invariant under the compact perturbation $2\eta^2$. This is easy to demonstrate, but before this, let us examine how the constraints we posed, that is ''compact'' and ''odd'', affect the matrix $2\eta^2$. Compact means that the matrix $\eta$ must contain finite numbers as elements and also odd means that $\eta^2$ must take the following form:
\begin{equation}\label{etasquareoddcccc}
\eta^2=\bigg{(}\begin{array}{ccc}
  0 & b \\
  -b & 0  \\
\end{array}\bigg{)},
\end{equation}
with $a=-b$. This can be true if the matrix $\eta$ is of the form:
\begin{equation}\label{etasquareoddcccceta}
\eta=\bigg{(}\begin{array}{ccc}
  \sqrt{b} & -\sqrt{b} \\
  \sqrt{b} & \sqrt{b}  \\
\end{array}\bigg{)}.
\end{equation}
Since the matrix $\eta^2$ is compact and odd, the following theorem holds true (see for example \cite{thaller} page 168, Theorem 5.28),
\begin{equation}\label{indperturbatrn1ffggg}
\mathrm{ind}_{t}(\mathcal{D}+C)=\mathrm{ind}_{t}\mathcal{D}
,\end{equation}
with $C$ a compact odd operator and $\mathcal{D}$ any trace class operator. In words, the regularized index of the operator $\mathcal{D}+C$ is equal to the index of the operator $\mathcal{D}$.
In our case, we are dealing with Fredholm operators and hence they are by definition trace-class. Now, owing to the theorem, the following relations hold true: 
\begin{align}\label{genkerrelationscompnew}
&-\Delta = \mathrm{dim}{\,}\mathrm{ker}\mathcal{D}_{u}^{\dag}\mathcal{D}_{u}-\mathrm{dim}{\,}\mathrm{ker}\mathcal{D}_{u}\mathcal{D}_{u}^{\dag}=
\\ \notag & \mathrm{dim}{\,}\mathrm{ker}(\mathcal{D}_{u}^{\dag}\mathcal{D}_{u}+2\eta^2)-\mathrm{dim}{\,}\mathrm{ker}(\mathcal{D}_{u}\mathcal{D}_{u}^{\dag}+2\eta^2)
\end{align}
Hence, the Witten index of the initial supersymmetric quantum mechanical system is invariant under the central charge extension of the system, for a real central charge and also with the relations (\ref{centrextendcc}), (\ref{etasquareoddcccc}) and (\ref{etasquareoddcccceta}) simultaneously holding true. This is easy to understand since both the operators $\mathcal{D}_{u}\mathcal{D}_{u}^{\dag}$ and $\mathcal{D}_{u}^{\dag}\mathcal{D}_{u}$ are trace-class (product of Fredholm--trace-class operators) and therefore the theorem applies to each of them.

\subsection{The Case of Equal Couplings for the Lepton-Down Quark Sector}

One common characteristic of many grand unified theories is that, when we make a renormalization group running of the couplings, the coupling constants become equal at some grand unification scale. In the grand unified theory under study, suppose that the couplings are unified at some grand unification scale $M_{gu}$. If the coupling constants of the lepton--down-quark sector become equal at some energy scale, this has a direct implication on the domain wall lepton--down-quark fermionic system, and particularly on the fermionic zero modes of the system. Actually, as we shall demonstrate, the two $N=2$, $d=1$ supersymmetries of each subsystem, combine to form an $N=4$, $d=1$ SUSY, with non-zero central charge. At the mass scale $M_{gu}$, we suppose that the coupling constants satisfy $g_e(M_{gu})=g_d(M_{gu})$. Recall that the kink solution $\phi_d$ couples in the same way to the down-quarks and to the charged leptons. In order to reveal the richer underlying SUSY QM structure of the system, we compute the following commutation and anti-commutation relations:
\begin{align}\label{commutatorsanticomm}
&\{{{\mathcal{Q}}_{d}},{{\mathcal{Q}}_{d}}^{\dag}\}=2\mathcal{H},{\,}\{{{\mathcal{Q}}_{e}},{{\mathcal{Q}}_{e}}^{\dag}\}=2\mathcal{H},{\,}\{{{\mathcal{Q}}_{e}},{{\mathcal{Q}}_{e}}\}=0,{\,}\{{{\mathcal{Q}}_{d}},{{\mathcal{Q}}_{d}}\}=0,{\,}{\,}\\
\notag & \{{{\mathcal{Q}}_{e}},{{\mathcal{Q}}_{d}}^{\dag}\}=\mathcal{Z},{\,}\{{{\mathcal{Q}}_{d}},{{\mathcal{Q}}_{e}}^{\dag}\}=\mathcal{Z},{\,}\\ \notag
&\{{{\mathcal{Q}}_{d}}^{\dag},{{\mathcal{Q}}_{d}}^{\dag}\}=0,\{{{\mathcal{Q}}_{e}}^{\dag},{{\mathcal{Q}}_{e}}^{\dag}\}=0,{\,}\{{{\mathcal{Q}}_{e}}^{\dag},{{\mathcal{Q}}_{d}}^{\dag}\}=0,{\,}\{{{\mathcal{Q}}_{e}},{{\mathcal{Q}}_{d}}\}=0{\,}\\
\notag
&[{{\mathcal{Q}}_{d}},{{\mathcal{Q}}_{e}}]=0,[{{\mathcal{Q}}_{e}}^{\dag},{{\mathcal{Q}}_{d}}^{\dag}]=0,{\,}[{{\mathcal{Q}}_{d}},{{\mathcal{Q}}_{d}}]=0{\,}[{{\mathcal{Q}}_{d}}^{\dag},{{\mathcal{Q}}_{d}}^{\dag}]=0,{\,}\\
\notag &
[{\mathcal{H}}_{d},{{\mathcal{Q}}_{d}}]=0,{\,}[{\mathcal{H}}_{d},{{\mathcal{Q}}_{d}}^{\dag}]=0,{\,}[{\mathcal{H}}_{e},{{\mathcal{Q}}_{e}}^{\dag}]=0,{\,}[{\mathcal{H}}_{e},{{\mathcal{Q}}_{e}}]=0,{\,}
\end{align}
with $\mathcal{Z}$:
\begin{equation}\label{zcentralcharge}
\mathcal{Z}=2\mathcal{H}
\end{equation}
Since at $M_{gu}$, we have $g_e(M_{gu})=g_d(M_{gu})$, the Hamiltonians of the down-quark--lepton system are equal, that is $\mathcal{H}={\mathcal{H}}_{e}={\mathcal{H}}_{d}$. The operator $\mathcal{Z}$ commutes with the supercharges ${{\mathcal{Q}}_{e}},{{\mathcal{Q}}_{d}}$, their conjugates ${{\mathcal{Q}}_{e}}^{\dag},{{\mathcal{Q}}_{d}}^{\dag}$ and
finally the Hamiltonians, $\mathcal{H}={\mathcal{H}}_{e}={\mathcal{H}}_{d}$.
The commutation and anti-commutation relations (\ref{commutatorsanticomm}) describe an $N=4$ supersymmetric quantum mechanics
algebra with central charge $\mathcal{Z}$. Recall that, for an $N=4$ algebra, the following relations hold true:
\begin{align}\label{n4algbe}
&\{Q_i,Q_j^{\dag}\}=2\delta_i^jH+Z_{ij},{\,}{\,}i=1,2 \\ \notag &
\{Q_i,Q_j\}=0,{\,}{\,}\{Q_i^{\dag},Q_j^{\dag}\}=0
\end{align}
The algebra described by relations (\ref{commutatorsanticomm}) actually possesses two central charges which
are equal and in particular the central charges are $Z_{12}=Z_{21}=\mathcal{Z}$. The fact the two $N=2$, $d=1$ SUSY QM algebras are enhanced at the unification scale $M_{gu}$ to an $N=4$, $d=1$, is a consequence of the fact that $g_e(M_{gu})=g_d(M_{gu})$ and also $\phi_e^{kink}(y)=\phi_d^{kink}(y)$. Notice that, the latter relation is a consequence of the fact that the domain wall solution couples in the same way the lepton and down-quark sector, as we already mentioned.

\noindent The $N=4$ supersymmetric algebra is particularly interesting since extended (with $N=4,6...$) supersymmetric quantum
mechanics models occur when $N=2$ and $N=1$ four dimensional Super-Yang Mills theories, are dimensionally reduced to $d=1$. It is intriguing that at the grand unification scale, the localized fermionic down quark-lepton system on the domain wall, has an underlying $N=4$, $d=1$ SUSY QM algebra. We could say that at that point, the SUSY QM algebraic structure of the system is enhanced from two $N=2$, $d=1$ to one $N=4$, $d=1$ SUSY QM algebra.

\subsection{A Comment on SUSY QM and Global Supersymmetry}

As we will demonstrate later on in this section, the spacetime manifold $M$ is locally a supermanifold and consequently one might wonder whether the SUSY QM algebra is connected to some global spacetime supersymmetry. A natural question to be asked, but the answer to it is no. 

\noindent Spacetime supersymmetric algebra, which is the well known graded Lie super-Poincare algebra in four dimensions, is four dimensional while the SUSY QM algebra is one dimensional. Moreover, spacetime supersymmetry in higher than one dimensions and
$d=1$ SUSY QM, are different for the simple reason that the $N=2$, $d=1$ SUSY QM supercharges
do not generate spacetime supersymmetry and thereby, SUSY QM does not directly relate the scalar and higher spin representations of the Poincare algebra in four dimensions, or alternatively put, SUSY QM does not relate fermions and bosons. The SUSY QM supercharges provide a $Z_2$ grading on the quantum states Hilbert space and additionally provides the quantum system with a set of transformations between the Witten parity eigenstates. As we shall evince later on in this section, this is the actual reason why the manifold $M$ is a graded manifold globally and not a supermanifold.

\noindent It is obvious that, the existence of a global supersymmetry does not necessarily entail the existence of a SUSY QM algebra on the quantum system that is supersymmetric. As we saw in the previous sections, the localized fermions do not belong to some supermultiplet since the initial system has no global supersymmetry, but nevertheless, the fermions belong to the quantum Hilbert space of $N=2$ SUSY QM. So it would be natural to think that a quantum system with initial global supersymmetry would have an underlying SUSY QM algebra with $N> 2$. This however is not necessarily true, as we exemplify in the following subsection.

\subsection{N=1 Global SUSY Domain Walls, Localized Fermions and $N=2$ SUSY QM}

In this section, we briefly study a model with global $N=1$ supersymmetry that admits domain wall solutions. As we shall demonstrate, the localized fermionic zero modes are connected to a $N=2$ SUSY QM algebra. Hence, although differently expected, due to the initial global supersymmetry, this system has an underlying $N=2$ SUSY QM structure too.

\noindent We shall examine the model studied in \cite{domainsusypaper}, and we also adopt the notation of that paper. We consider a supersymmetric model with two chiral superfields,
\begin{equation}\label{phifield}
\Phi_i(x)=\phi_i(x)+\sqrt{2}\theta \psi_i (x)+\theta^2 F_i(x)
\end{equation}
with $F_i$ the auxiliary fields. The $N=1$ supersymmetric action is,
\begin{align}\label{actionssusyn1}
S=\int \mathrm{d}x^4\mathrm{d}\theta^2\mathrm{d}\bar{\theta}^2\bar{\Phi_i^*}(x)\Phi_i(x)+\int \mathrm{d}x^4\mathrm{d}\theta^2W(\Phi)+\int \mathrm{d}x^4\mathrm{d}\bar{\theta}^2 W^*(\Phi^*)
\end{align}
with the superpotential being equal to:
\begin{equation}\label{superpotential}
W(\Phi)=\frac{1}{2}\lambda \Phi_2(\Phi_1^2-v^2)
\end{equation}
The fermionic equations of motion for general $\phi_{1,2}$ solutions are equal to (see \cite{domainsusypaper} for more details):
\begin{align}\label{fermioniceqnsmotionsn1}
&\gamma^{\mu}\partial_{\mu}\Psi_1+2\lambda\Big{[}(\phi_1P_L+\phi_1^*P_R)\Psi_2+(\phi_2P_L+\phi_2^*P_R)\Psi_1\Big{]}
\\ \notag & \gamma^{\mu}\partial_{\mu}\Psi_2+2\lambda (\phi_1P_L+\phi_1^*P_R)\Psi_1=0
\end{align}
with $\Psi_i$ Majorana spinors. The fermion fields become effectively massless in the core of the domain wall, allowing fermionic zero modes to form. In the domain wall background $\phi_2=0$  and $\phi_1=\phi_w(x)$ (with $\phi_w$ a real function) the fermionic equations of motion become:
\begin{align}\label{fermioniceqnsmotionn1susy}
&\gamma^{\mu}\partial_{\mu}\Psi_1+2\lambda\phi_w\Psi_2=0
\\ \notag & \gamma^{\mu}\partial_{\mu}\Psi_2+2\lambda \phi_w\Psi_1=0
\end{align}
Equations (\ref{fermioniceqnsmotionn1susy}) have the following localized solutions: 
\begin{equation}\label{locsoln1susylocal}
\Psi_1=\tau e^{-2\lambda \int_0^x\phi_w(x')\mathrm{d}x'},{\,}{\,}{\,}\Psi_2=\gamma^1\Psi_1(x)
\end{equation}
with $\tau$ an arbitrary constant Majorana spinor. These solutions are associated to effectively massless fermions trapped within the core of the domain wall travelling in the $+z$ and $-z$ directions and hence, the domain wall is superconducting. Particularly, we have fermionic domain wall superconductivity since the domain wall can support fermionic charge and current. From equations (\ref{fermioniceqnsmotionsn1}) we can construct an unbroken $N=2$ SUSY QM, with supercharge $Q_{s}$:
\begin{equation}\label{s7supcghagemixsusyg}
\mathcal{Q}_{s}=\bigg{(}\begin{array}{ccc}
  0 & \mathcal{D}_{s} \\
  0 & 0  \\
\end{array}\bigg{)},{\,}{\,}{\,}\mathcal{Q}_{s}=\bigg{(}\begin{array}{ccc}
  0 &  \mathcal{D}_{s}^{\dag} \\
 0 & 0  \\
\end{array}\bigg{)}
\end{equation}
and quantum Hamiltonian:
\begin{equation}\label{s11fgghhfsusyg}
\mathcal{H}_{s}=\bigg{(}\begin{array}{ccc}
 \mathcal{D}_{s}\mathcal{D}_{s}^{\dag} & 0 \\
  0 & \mathcal{D}_{s}^{\dag}\mathcal{D}_{s}  \\
\end{array}\bigg{)}
\end{equation}
In the above equations, the operator $\mathcal{D}_s$ is equal to:
\begin{equation}\label{susyqmrn567msusyg}
\mathcal{D}_{s}=\left(%
\begin{array}{cc}
\gamma^{\mu}\partial_{\mu} & 2\lambda \phi_w
 \\ 2\lambda \phi_w & \gamma^{\mu}\partial_{\mu}\\
\end{array}%
\right)
\end{equation}
Following the same line of research as in the previous sections, we can easily establish the result that SUSY QM is unbroken for this system too. Notice that the system has $N=2$ SUSY QM although the initial system has an $N=1$ global supersymmetry. Hence, this could suffice as a counter example to the argument that the existence of an initial global supersymmetry and hence the existence of supercharges, has an underlying $N>2$ extended supersymmetry quantum mechanics algebra.

\section{Some Implications of the SUSY QM Algebra on the Quantum System}

\subsection{A Global R-symmetry of the Quantum Hilbert Space}

The existence of an $N=2$ SUSY QM algebra, has a direct implication on the Hilbert space quantum states. As we shall now demonstrate, the quantum algebra is invariant under a global $U(1)$ symmetry. We perform the following transformation on the supercharges ${\mathcal{Q}}_{u}$ and ${\mathcal{Q}}^{\dag}_{u}$:
\begin{align}\label{transformationu1}
& {\mathcal{Q}}_{u}^{'}=e^{-ia}{\mathcal{Q}}_{u}, {\,}{\,}{\,}{\,}{\,}{\,}{\,}{\,}
{\,}{\,}{\mathcal{Q}}^{'\dag}_{u}=e^{ia}{\mathcal{Q}}^{\dag}_{u}
.\end{align}
The quantum Hamiltonian of the system is invariant under this global transformation, but the quantum states of the system are transformed accordingly. Particularly, recall that there is a $Z_2$ grading on the total Hilbert space, and so the graded quantum states $\psi^{+}_{M}$ $\in$ $\mathcal{H}^{+}_{M}$ and
$\psi^{-}_{M}$ $\in$ $\mathcal{H}^{-}_{M}$, namely the even and odd states respectively, are transformed under the $U(1)$ transformation, as follows:
\begin{equation}
\psi^{'+}_{M}=e^{-i\beta_{+}}\psi^{+}_{M},
{\,}{\,}{\,}{\,}{\,}{\,}{\,}{\,}
{\,}{\,}\psi^{'-}_{M}=e^{-i\beta_{-}}\psi^{-}_{M}
.\end{equation}
Of course, the parameters $\beta_{+}$ and $\beta_{-}$ are global
parameters that are connected to $a$ as $a=\beta_{+}-\beta_{-}$.

\subsection{A Global two term Spin Complex Structure}

The $N=2$ SUSY QM algebra of the quantum system of the localized fermions on a domain wall, provides the quantum system with an additional geometric structure. Particularly, the fermionic system in terms of the supercharges has an inherent spin complex structure, which we now present. As we already mentioned, the Witten parity $\mathcal{W}$, makes the Hilbert space of the SUSY QM mechanics algebra $Z_2$-graded, so that $\mathcal{H}(M)=\mathcal{H}^+(M)\oplus \mathcal{H}^-(M)$, with $M$ denoting the corresponding spacetime manifold, on which the fermions are sections of the corresponding spin fibre bundle. The vectors $|\psi^{+}\rangle$ $\in$ $\mathcal{H}^+(M)$ and $|\psi^{-}\rangle$ $\in$ $\mathcal{H}^-(M)$, are equal to, 
\begin{align}\label{phi5gdfghfdh}
&|\psi^{+}\rangle =\left(%
\begin{array}{c}
  |\phi ^{+} \rangle \\
  0 \\
\end{array}%
\right),{\,}{\,}{\,}\in{\,}{\,}\mathcal{H}^+(M)
\\ \notag &
|\psi^{-}\rangle =\left(%
\begin{array}{c}
  0 \\
  |\phi ^{-} \rangle \\
\end{array}%
\right),{\,}{\,}{\,}\in{\,}{\,}\mathcal{H}^-(M),
\end{align}
with $|\phi^{\pm}\rangle$, the vectors corresponding to the matrices $\mathcal{D}_i$, defined in the previous sections. When the supercharges ${\mathcal{Q}_u}$, ${\mathcal{Q}_u}^{\dag}$ act on the vectors $|\psi^{\pm}\rangle$, we get:
\begin{align}\label{wittyuhjhjhhtyi2}
&\bigg{(}\begin{array}{ccc}
  0 & \mathcal{D}_i \\
  0 & 0  \\
\end{array}\bigg{)}\left(%
\begin{array}{c}
   0\\
  |\phi ^{-} \rangle \\
\end{array}\right)=\left(%
\begin{array}{c}
  \mathcal{D}_i|\phi ^{-} \rangle \\
  0 \\
\end{array}\right),{\,}{\,}{\,}\in{\,}{\,}\mathcal{H}^+(M)
\\ \notag & \bigg{(}\begin{array}{ccc}
  0 & 0 \\
  \mathcal{D}_i^{\dag} & 0  \\
\end{array}\bigg{)}\left(%
\begin{array}{c}
  |\phi ^{+} \rangle \\
  0 \\
\end{array}\right)=\left(%
\begin{array}{c}
   0\\
  \mathcal{D}_i^{\dag}|\phi ^{+} \rangle \\
\end{array}\right),{\,}{\,}{\,}\in{\,}{\,}\mathcal{H}^-(M)
.\end{align}
Consequently, the supercharges generate the following two maps:
\begin{align}\label{mapsscharge}
&{\mathcal{Q}_u}:\mathcal{H}^-(M)\rightarrow \mathcal{H}^+(M)
\\ \notag & {\mathcal{Q}_u}^{\dag}:\mathcal{H}^+(M)\rightarrow \mathcal{H}^-(M)
.\end{align}
With these two maps, a two-term spin complex is constructed, which has the following form:
$$\harrowlength=40pt \varrowlength=.618\harrowlength
\sarrowlength=\harrowlength
\mathcal{H}^+(M)\commdiag{\mapright^{{\mathcal{Q}_u}^{\dag}} \cr \mapleft_{{\mathcal{Q}_u}}}\mathcal{H}^-(M)$$
The index of this two-term spin complex is identical to the Fredholm index of the operator $\mathcal{D}_i$.

\subsection{Some Local Geometric Implications of the SUSY QM Algebra on the Spacetime Fibre Bundle Structure}

The $N=2$ SUSY QM algebra, has some local geometric implications on the fibre bundle structure of the spacetime $M$. As we show, due to the $N=2$ SUSY QM algebra, the spacetime manifold $M$ is locally rendered a supermanifold. The supercharge of the SUSY QM algebra is actually the local superconnection on this supermanifold, and the square of the supercharge is the corresponding curvature. For an important stream of papers and textbooks discussing the mathematical issues we will use, see \cite{graded1,Jost}.

\noindent The localized fermions on the domain walls, in the spacetime $M$, are actually sections of the $U(1)$-twisted fibre bundle $P\times S \otimes U(1)$. In this spacetime, $S$ denotes the reducible representation of the Spin group $Spin(4)$, and $P$ is the double cover of the principal $SO(4)$ bundle on the tangent manifold $TM$. A $Z_2$ grading on a vector space $E$, is performed by decomposing the vector space in the following way:
\begin{equation}\label{z2grad}
E=E_+\oplus E_{-}
\end{equation}
In addition, a $Z_2$-grading of an algebra $A$ to even and odd elements, $A=A_+\oplus A_{-}$, is done in such a way so that the following hold true: 
\begin{equation}\label{amodule}
A_+\cdot E_+\subset E_+,{\,}{\,}A_+\cdot E_-\subset
E_-,{\,}{\,}A_-\cdot E_+\subset E_-,{\,}{\,}A_-\cdot E_-\subset
E_+,
\end{equation}
The algebra $A$ is called a $Z_2$-graded algebra. We denote with $W$ an involution which belongs to the set of endomorphisms of $E$, which we denote $\mathrm{End} (E)$ (actually $W$ is the Witten parity operator). The involution $\mathcal{W}=\pm1$ acts on the vectors of $E$, in the following way:
\begin{equation}\label{befoend}
\mathcal{W}(a+b)=a-b,{\,}{\,}{\,}\forall{\,}{\,}a{\,}\in{\,}E_+,{\,}{\,}\mathrm{and}{\,}{\,}\forall{\,}{\,}b{\,}\in{\,}E_-.
\end{equation} 
The involution $\mathcal{W}$, provides the algebra $\mathrm{End} (E)$ with a $Z_2$-grading. So it renders it a $Z_2$-algebra. These very general considerations have a direct application to the case of the $N=2$ SUSY QM quantum Hilbert space $\mathcal{H}$ that we study. Particularly, the elements of $\mathrm{End} (\mathcal{H})$ are matrices of the form:
 \begin{align}\label{evenodd}
&\bigg{(}\begin{array}{ccc}
  0 & g_1 \\
  g_2 & 0  \\
\end{array}\bigg{)},{\,}{\,}{\,}\mathrm{odd}{\,}{\,}{\,}\mathrm{elements}
\\ \notag &
\bigg{(}\begin{array}{ccc}
 g_1 & 0 \\
  0 & g_2  \\
\end{array}\bigg{)},{\,}{\,}{\,}\mathrm{even}{\,}{\,}{\,}\mathrm{elements},
\end{align}
with $g_1,g_2$ generally complex numbers. As we have seen, the involution $\mathcal{W}$, generates the $Z_2$ graded vector space
$\mathcal{H}=\mathcal{H}^+\oplus \mathcal{H}^-$, and the subspace
$\mathcal{H}^+$ contains $\mathcal{W}$-even vectors while $\mathcal{H}^-$, $\mathcal{W}$-odd vectors.
Hence, an additional algebraic structure is defined on the manifold $M$, which is owing to this $Z_2$-grading. We denote this additional $Z_2$-graded algebra $\mathcal{A}$, with
$\mathcal{A}=\mathcal{A}^+\oplus \mathcal{A}^-$. In the case at hand, the algebra $\mathcal{A}$ is a total rank two sheaf of
$Z_2$-graded commutative $R$-algebras. Hence $M$
becomes a graded manifold $(M,\mathcal{A})$, but not a supermanifold (at least globally).

\noindent The endomorphism $\mathcal{W}$, $\mathcal{W}:\mathcal{H}\rightarrow
\mathcal{H}$ is contained in the sheaf $\mathcal{A}$, and hence $\mathrm{End}(\mathcal{H})\subseteq \mathcal{A}$. The sheaf $\mathcal{A}$ is called the structure sheaf of the graded manifold
$(M,\mathcal{A})$, while the manifold $M$ is the body of
$(M,\mathcal{A})$. Locally, the structure sheaf $\mathcal{A}$ is isomorphic to the sheaf
$C^{\infty}(U)\otimes\wedge R^m$ of the exterior affine vector bundle
$\wedge \mathcal{H_E}^*=U\times \wedge R^m$. The affine vector bundle $\mathcal{H_E}$
has fiber the vector space $\mathcal{H}$ and $U$ is an open set of the manifold $M$. The structure sheaf $\mathcal{A}=C^{\infty}(U)\otimes\wedge \mathcal{H}$, is
isomorphic to the sheaf of sections of the exterior vector bundle
$\wedge \mathcal{H_E}^*=R\oplus
(\oplus^{m}_{k=1}\wedge^k)\mathcal{H_E}^*$. Let us see the local geometric implications of the aforementioned sheaf structure. The sections of the fibre bundle $TM^*\otimes\mathcal{H}$ are identical to the sections of the fermionic bundle $P\times S \otimes U(1)$, related to the SUSY QM algebra. A local superconnection, denoted $\mathcal{S}$, is an 1-form with values in $\mathrm{End} (E)$. To put this differently, a local superconnection is a section of  $TM^*\otimes \wedge \mathcal{H_E}^*\otimes
\mathcal{H_E}$. The corresponding curvature of the superconnection, which we denote $\mathcal{C}$, is an $\mathrm{End} (E)$-valued 2-form on $M$, which satisfies:
\begin{equation}\label{}
\mathcal{C}=\mathcal{S}^2
\end{equation}
Consequently, locally on $M$, the superconnection is a section of the fibre bundle $TM^*\otimes \mathrm{End} (E)^{\mathrm{odd}}$. The latter contains the odd elements of $\mathrm{End} (E)$. Therefore, at least locally (and locally means at an infinitesimally small open neighborhood of a point $x$ $\in$ $M$), the supercharge of the SUSY QM algebra is the superconnection, that is $\mathcal{S}={\mathcal{Q}_u}$, and hence, the curvature of the supermanifold is locally $\mathcal{C}={\mathcal{Q}}^2_u$. 

\noindent Recapitulating, we can say that the $N=2$ SUSY QM structure locally renders the manifold $M$ a supermanifold, with superconnection ${\mathcal{Q}_u}$ and curvature ${\mathcal{Q}}^2_u$. Moreover, the manifold $M$ is globally a graded manifold $(M,\mathcal{A})$ with structure sheaf $\mathcal{A}$ and body $(M,\mathcal{A})$.

\section*{Concluding Remarks}

In this article we studied a particular attribute of the localized fermions on superconducting domain walls. Specifically, we demonstrated that an unbroken $N=2$, $d=1$ SUSY QM algebra underlies each fermion family localized on the domain wall. The domain wall, originating from a specific grand unified theory, is coupled to three fermion families, namely the up-quark, down-quark and to charged leptons. In addition, the domain wall couples to the down-quark and the leptons in the same way. As we substantiated, the three independent $N=2$ SUSY QM algebras can be combined to higher reducible representations of $N=2$, $d=1$ supersymmetry, with or without central charge. Moreover, in some cases, we were able to construct SUSY QM algebras that are invariant under certain duality transformations, with the latter having to do with the couplings of the quark and lepton fields. In the particular case that the couplings of the lepton and down-quark fields become equal, something that is possible at the coupling unification scale, the two $N=2$, $d=1$ SUSY QM algebras combine to form an $N=4$, $d=1$ SUSY QM algebra with central charge. Furthermore, we studied an $N=1$ global supersymmetric domain wall model, and we established the result that an unbroken $N=2$, $d=1$ supersymmetry underlies this fermionic system too. After that, we presented some extra characteristics of the localized fermions on the domain wall, owing to the existence of the SUSY QM algebras. Some geometrical implications were presented too.

Before closing, let us discuss one important issue, having to do with $d=4$ global supersymmetry and $d=1$ supersymmetry. In the cases we studied, and particularly in the grand unified model case, there was no initial supersymmetry in the system, but in the end we found unbroken $N=2$, $d=1$ supersymmetry. It stands to reason that it would be expected to find an extended supersymmetry structure, if an initial global supersymmetry exists in the system. This however is not true as we evinced explicitly, in the case of an $N=1$ model. So in conclusion, there is no direct connection between global spacetime supersymmetry and SUSY QM algebras, at least in the context we used in this article. However, when the system possesses an initial $N=2$, $d=3$ global supersymmetry, an extended SUSY QM algebra underlies the system, see for example \cite{oikonomoucs}. This is not however a general rule, but just an observation made for a particular system.

\end{document}